\documentclass[11pt,rotating]{article}
\input{epsf}

\usepackage{times}
\usepackage{graphicx}
\usepackage{fancyhdr}
\usepackage{amssymb}
\usepackage{amsmath}
\usepackage{fancybox}

\date{}

\begin{document}

\title{Universality issues in surface kinetic roughening of thin solid films}
\author{
Rodolfo Cuerno\footnote{Depto.\ de Matem\'aticas and GISC, Universidad
Carlos III de Madrid, Avenida de la Universidad 30, 28911 Legan\'es, Spain.
{\tt cuerno@math.uc3m.es}}
\and Luis V\'azquez\footnote{Instituto de Ciencia de Materiales de Madrid,
28049 Cantoblanco (Madrid), Spain. {\tt lvb@icmm.csic.es}}}


\maketitle 

\begin{abstract}
Since publication of the main contributions on the theory of kinetic 
roughening more than fifteen years ago, many works have been reported on 
surface growth or
erosion that employ the framework of dynamic scaling. This interest was mainly 
due to the predicted existence of just a few universality classes to 
describe the statistical properties of the morphology of growing surfaces and 
interfaces that appear in a wide range of physical systems. Nowadays, this 
prediction seems to be inaccurate. This situation has caused a clear 
detriment of these studies in spite of the undeniable existence of kinetic 
roughening in many different real systems, and without a clear understanding 
of the reasons behind the mismatch between theoretical expectations and 
experimental observations. In this chapter we aim to explore existing
problems and shortcomings of both the theoretical and experimental approaches, 
focusing mainly on growth of thin solid films. Our analysis suggests 
that the theoretical framework as yet is not complete, while more systematic 
and consistent experiments need to be performed. Once these issues are taken 
into account, a more consistent and useful theory of kinetic roughening might 
develop.
\end{abstract}

\vspace*{-12cm}
\begin{flushleft}
{\scriptsize To appear in \textit{
Advances in Condensed Matter and Statistical Physics} \\ \vspace*{-0.1cm}
E.\ Korutcheva and R.\ Cuerno, editors (Nova Science Publishers, New York, 2004)}
\end{flushleft}
\vspace*{12cm}

\section{Introduction}

During the last few years the research community has devoted a great
effort to the understanding of a wide variety of interface growth
phenomena under the framework of dynamic scaling concepts. These
phenomena range from thin solid film growth and erosion processes to the
growth dynamics of systems that were beyond the realm of traditional
Statistical Mechanics or Condensed Matter Physics, such as tumors,
bacterial colonies, etc. \cite{alb,krug,meakin}. The appeal to describe
the interface fluctuations observed in such diverse experimental
domains is due to the simplicity of the original framework within
which the various observations were described, namely the scaling
Ansatz due to Family and Vicsek \cite{fv}.  Thus, the {\em surface
or interface width} or {\em roughness}, $W(t)$, defined as the rms
deviation of the height field $h(\mathbf{r},t)$ around its mean value
$\bar{h}(t)$, is seen to increase with time as a power law $W(t) \sim 
t^{\beta}$ for as long as the lateral correlation length 
$\xi(t) \sim t^{1/z}$ is smaller than the system size, $L$. For longer times
the roughness becomes a constant $W \sim L^{\alpha}$.  The scaling
exponents $\alpha$, $\beta$, and $1/z$ quantify thus, respectively,
the spatial correlations of interface features, the temporal evolution
of the interface roughness, and the coarsening process of the
characteristic lateral correlation length of the interface, for a
system displaying scale invariance in time and space.  This behavior
is a natural extension to a growing surface or interface, of the {\em
dynamic} scaling hypothesis for equilibrium dynamics of critical
systems \cite{hh}. Actually, the high temperature phase of the
discrete Gaussian model, intensively studied in the context of the
equilibrium roughening transition of a metallic surface (see e.g.\
\cite{weeks,nozieres,lapujolade}), already provided a well known
example of a surface whose equilibrium dynamics fits completely with
the FV Ansatz.\footnote{In the kinetic roughening context, the
corresponding universality class is termed Edwards-Wilkinson (EW)
\cite{ew}, making reference to the description by the corresponding
Langevin equation.}  From the theoretical point of view, this fact
encouraged the modeling of surface/interface growth by the use of
discrete models of the Monte Carlo type, or by stochastic differential
equations similar, as in the discrete Gaussian model,
to, e.g.\ models A and B of critical dynamics, in the classification
of Ref.\ \cite{hh}. These approaches were simplified by the implied
existence of {\em universality} and universality classes,
identified by the values of the critical exponents $\alpha$, and
$z$.\footnote{Substitution of $L \to \xi(t) \sim t^{1/z}$ in the
saturation value of the roughness leads to the scaling relation
$\beta=\alpha/z$, which indeed occurs for a flat initial condition.}
Moreover, the main source of phenomenology supporting the FV Ansatz
was in the field of {\em fractal growth} \cite{vicsek}. Indeed the
roughness exponent $\alpha$ coincides with the so-called Hurst
exponent of the surface, seen as a {\em self-affine} fractal, and is
thus related to the, say box-counting fractal dimension $D_B$, as
\cite{meakin,vicsek} $\alpha = d_E - D_B$, where $d_E$ is the
embedding Euclidean dimension.\footnote{We will employ $d$ for the
dimension of the substrate ``on top'' of which growth occurs along an
additional dimension, thus $d_E = d+1$. E.g.\ a one-dimensional random
walk is a rough ``surface'' for which $d_E = d + 1 = 2$.}  Remarkable
examples of non-equilibrium surfaces fulfilling the FV Ansatz were
found to be the active zones of the {\em diffusion-limited
aggregation} (DLA) or Eden models, much studied in the wider
context of {\em pattern formation} outside of equilibrium \cite{ball}.
Thus, by the mid-1980's enough theoretical (analytical and
simulational) and some experimental evidence strongly conveyed the
vision that the large scale properties of rough surfaces grown in
equilibrium and far from equilibrium processes should fall into a few
universality classes. One of these would be provided by the asymptotic
properties of the celebrated Kardar-Parisi-Zhang (KPZ) equation
\cite{kpz}, itself directly related to other non-equilibrium systems,
such as fluid turbulence or directed polymers in random media
\cite{krug-spohn,hhz}.  This appealing perspective triggered the
interest of the scientific community. Thus, scaling analysis was
applied to a large variety of theoretical and experimental systems
(see in particular the compilation of experiments reviewed in
\cite{alb,meakin,kp}). Note that, given the interpretation of the
scaling exponents, a natural procedure in an experiment is to, first,
evaluate these and then, within the universality assumption, to try
contrast them with those theoretically predicted in order to fully
understand the mechanisms ruling the growth process. This deeper
knowledge of the physical mechanisms governing the interface growth
dynamics can have implications in many fields. For instance, in the
growth of thin solid films, it can contribute to a better control of the
deposition conditions in order to produce films with technological
applications (electronic, electro-optical, catalytic, etc.) including
fabrication of nanostructures \cite{facsko}. Also for instance, in the
growth of tumors \cite{bru}, this type of identification has led to
the conclusion that actual cell duplication (growth) takes place on
the surface, rather than, as traditionally assumed, in the bulk, which
might lead to appropriate techniques in order to, say, halt tumor
growth.  However, when trying to apply the above framework to
understand the fluctuation properties of actual surfaces, in
particular in the context of thin solid film production, the experimental
evidence was far less compelling than implied by such a simple,
unifying theoretical picture.  Unfortunately, as noted in \cite{lee},
``while many experiments on surface roughening during film growth have
been interpreted in the context of the dynamic scaling hypothesis, no
consensus has emerged concerning the relationship between roughness
and growth exponents appropriate to a given growth process''. This
situation led M.\ Kardar to state in \cite{kardar}, that ``despite its
ubiquitous occurrence in theory and simulations, experimental
confirmation of dynamic scaling has been [relatively, this is ours]
scarce. In some cases where such scaling has been observed, the
exponents are different from those expected''.
In fact, there do exist many experiments in which the surface
fluctuations exhibit scale invariance. However, the spread in the
values of the ensuing exponents (see e.g.\ Table 2 in \cite{kp})
questions the very existence of universality in these phenomena, not
to mention the difficulty quoted in attributing a given set of
experimental exponents to a specific physical mechanism.

Nowadays, seven years after publication of the above quotations, 
the situation seems to be quite similar
to that described in the reviews cited above. It is indeed true that
some systems have been analyzed, and fully and consistently assigned
to some of the main theoretical universality classes. Examples are
found in experiments on paper burning \cite{finn2} for the KPZ class
in $d_E=2$, sputtered Fe/Au multilayers \cite{paniago}, silica films
grown by chemical vapor deposition \cite{us-cvd}, or growth/etching of
epitaxial GaAs films \cite{ballestad}, for the KPZ class in $d_E = 3$,
or sedimentation of silica nanospheres \cite{lvb} for the EW class in
$d_E = 3$.  However, many other systems have not been explained until now,
and basically remain just phenomenologically described by a set of
experimental exponents. Even in the study of highly idealized models of 
surface growth by Molecular Beam Epitaxy (MBE), such as those of Wolf-Villain 
(WV) \cite{wv} and Das Sarma-Tamborenea \cite{dt} (DT) ---see below---, 
the very notion of universality can be questioned \cite{dspt}. 
Thus we are forced into the question whether a consistent physical theory 
of kinetic roughening can be formulated, and with what use. 
Given the undeniable experimental evidence on surface kinetic roughening,
we believe the answer is positive, but the development of the theoretical
framework and tools are as yet incomplete. Moreover, the approach has to 
be inverted to some extent: rather than postulating universality and trying 
to fit observations to ``idealized'' universality classes, the generalization 
to these need come after detailed study of well defined systems. On the way, 
this process will also provide new developments in the field of generic
models of non-equilibrium systems.

In this chapter we investigate the reasons for the mismatch between
theory and experiment. Our aim is to assess the current status of 
kinetic roughening as a consistent physical theory, through the 
comparison between the ``ideal'' universality classes, as represented e.g.\ 
by the continuum stochastic equations of EW, KPZ, etc., and
experimental observations. We will thus evaluate the practical relevance 
of the universality concept ---including its very definition in some 
instances---, from the point of view of its pertinence to and of the
the physical information it provides on the experimental systems. 
We will not dwell on more fundamental questions such as the origin of 
scale invariance properties in these systems, see e.g.\ 
\cite{kadanoff}, neither on the generic properties of the 
representatives of the ``ideal'' universality classes as instances of 
non-equilibrium systems with strong fluctuations.
As mentioned in the previous paragraph, 
our approach is that these questions might 
obtain some answers by generalizing results obtained after 
close comparison between theory and experiments in well delimited
physical instances. Interestingly, we will see that 
progress can be made here by strengthening and exploiting the  
connection with those fields in which the subject originated in the
first place, namely, those of {\em pattern formation} and 
{\em fractal geometry}.
We will restrict ourselves almost exclusively to the relatively reduced 
domain of thin solid film growth, and those immediately linked to it, while
we expect that much of our discussion will carry over to other domains
of applicability of the framework provided by dynamic scaling Ansatz.
Most notably, we will {\em not} consider the specific case of kinetic
roughening in media with {\em quenched disorder}. These are reviewed
elsewhere for the case of fluid imbibition, see Ref.\ \cite{finn}, in
a spirit that is somehow akin to that of the present chapter. Moreover, we 
will restrict our attention, both in the discussion of theoretical and
experimental approaches, mostly to those issues which we consider that
are not properly understood, or that are promising lines of research
with the goal of developing a more consistent picture of the field.

\section{Status of the theory}

\subsection{``Ideal" universality classes} \label{sec_ideal}

In this section, we briefly recall the main universality classes
relevant for kinetic roughening. As stated above, we focus on the case
in which the surface or interface is subject to time dependent noise
\cite{ojalvo}. In typical applications, these fluctuations arise in
those of a driving flux (of, say, aggregating units, atoms or
molecules) acting on the system. This is a convenient way to represent
the nature of the noise, but it does not by any means imply that its
amplitude is directly the square root of the average
external flux.\footnote{E.g., under the assumption of
Poisson statistics ({\em shot noise}).} For instance, in studies of growth by MBE
\cite{pierre-louis} or by electrochemical or chemical vapor deposition
(ECD, CVD, respectively) \cite{mario} the noise term appearing in the
Langevin equation for the interface is seen to be rather more involved
than that.  However, and this can never be overemphasized, the
universal behavior applies to asymptotic properties, well beyond all
existing transients (induced by, e.g.\ physical instabilities acting
on the system) and crossovers (due to competition among various
physical mechanisms, each of which is dominant for a different range
in time and space). For the type of systems we are considering, the
asymptotic properties are adequately described by equations featuring
additive noise, which is Gaussian and uncorrelated in time and space.

Below follows a summary of the current consensus \cite{ds} on the universality 
classes for kinetic roughening. For each of these, the representative 
continuum equation for the surface height is provided. Again, this has to be
understood in the renormalization group (RG) sense, namely, the fact that
a given experimental system belongs to one of these universality classes
does {\em not necessarily} mean that the surface height ---seen at the spatial
and temporal scales resolved in the experiment--- follows ``in detail'' the 
corresponding Langevin equation. Rather, it means that both systems display 
the same statistical properties.\footnote{In that sense, the continuum 
equation can be thought of as an {\em effective} description of the physical 
system at appropriate {\em coarse-grained} space and time scales.} 
A natural and standard classification of the universality classes 
considers the existence of conservation laws on the surface height in the 
representative equation. Thus, we have:
\begin{enumerate}
\item Non conserved systems: If the height evolves in time in the absence
of a conservation law on the mass of aggregating units at the interface
(such is the case for, e.g., ECD and CVD), for instance if bulk vacancies are
significative, or conformal growth occurs \cite{us-cvd}, then the expected
asymptotic behavior is that of the KPZ equation
\begin{equation}
\frac{\partial h}{\partial t} = \nu \nabla^2 h + \frac{\lambda}{2} (\nabla h)^2
+ \eta .
\label{kpz}
\end{equation}
In this equation, the first two terms on the right hand side are, 
to lowest order, 
{\em representing} respectively surface tension effects ($\nu$ is a positive 
constant) and growth along the local normal direction with a constant average
rate, $\lambda$. In this and subsequent interface equations, 
$\eta(\mathbf{r},t)$ is the noise term alluded to in the previous paragraph. 
The scaling exponents characterizing this universality class are exact in 
$d_E=2$, $\alpha = 1/2$, and $z=3/2$ ($\beta=1/3$), and approximately equal 
to $\alpha = 0.39$ and $z=1.61$ ($\beta=0.24$) in $d_E=3$ \cite{marinari}. 
This universality class is exceedingly important in the theory of 
non-equilibrium processes.\footnote{See e.g.\ the chapter by 
M.\ A.\ Mu\~noz in this book.} However, as mentioned above there are very 
few experimental realizations of this scaling behavior. 

\item Conserved systems:
The other three main universality classes in kinetic roughening refer
to systems in which, apart from the noise term, the representative
Langevin equation has the shape of a continuity equation, reflecting 
a conservation law on the mass of aggregating units {\em at} the surface.
Such kind of symmetry may arise dynamically in an effective way, or be
explicitly manifest in the physical mechanisms leading to growth.
Thus we have:

\begin{enumerate}

\item EW universality class: 
There are surfaces whose dynamics can be asymptotically described 
by an evaporation-condensation effect (combined with fluctuations)
of the Gibbs-Thompson type. This is the case, for instance, in CVD 
experiments with non-negligible vapor pressure. Note that
in this case, apart from the external flux, the surface evolves as 
around an equilibrium state. The EW Langevin equation is thus a linearization
of (\ref{kpz}):
\begin{equation}
\frac{\partial h}{\partial t} = \nu \nabla^2 h + \eta .
\label{ew}
\end{equation}
For any value of $d_E = d+1$, the critical exponents characterizing this 
universality class are $\alpha= (2-d)/2$ and $z=2$ [$\beta= (2-d)/4$]. 
These were already
familiar from the studies of the equilibrium roughening transition,
as mentioned in the Introduction, Eq.\ (\ref{ew}) being simply model A
for the Ginzburg-Landau functional in the Gaussian approximation.

\item Linear MBE universality class:
In the same way as Eq.\ (\ref{ew}) can be thought of, in the simplest 
representation, as the equilibrium dynamics of a surface minimizing its 
area \cite{weeks,nozieres}, an important class of rough surfaces can be 
described effectively as minimizing their mean curvature.\footnote{In analogy
with the discrete Gaussian model, there is also a roughening transition
for these systems, described through the discrete {\em Laplacian} model
\cite{nelson}.} The Langevin equation representing this universality
class reads
\begin{equation}
\frac{\partial h}{\partial t} = - \nu \nabla^4 h + \eta ,
\label{lmbe}
\end{equation}
Again (as reflected in the previous equation being linear) the
scaling exponents are exact in this case for any value of $d_E$, namely, 
$\alpha = (4-d)/2$ and $z=4$ [$\beta= (4-d)/8$]. 
Although the thermodynamic limit of
a system like (\ref{lmbe}) is {\em not} well defined for $d_E= 2$, $3$, this does not
prevent real (and finite) systems to display the referred scaling behavior
\cite{jeff}.

\item Non-linear MBE universality class:
To some extent in analogy with the relationship between the KPZ and EW 
equations, much work has been devoted to elucidating the appropriate
non-linear extension of (\ref{lmbe}) leading to a well-defined 
universality class. It is represented by the Langevin equation
\begin{equation}
\frac{\partial h}{\partial t} = - \nu \nabla^4 h + \lambda \nabla^2 
(\nabla h)^2 + \eta .
\label{nlmbe}
\end{equation}
In this case the critical exponents are close to \cite{janssen}
$\alpha = 1$, $z=3$ [$\beta= 1/3$] in $d_E = 2$, and $\alpha = 2/3$ 
and $z=10/3$ [$\beta= 1/5$] in $d_E = 3$.
\end{enumerate}
\end{enumerate}
Although the above universality classes are those which have focused most 
work in the field of kinetic roughening, the above list is possibly
far from complete. Actually, in a certain way there is a {\em continuum 
of universality classes}! Some not covered in the previous list arise
in physical mechanisms which cannot be ruled out {\em a priori} when 
considering applications to real data:
\begin{itemize}

\item Noise properties: It is well known theoretically that the introduction 
of particular features in the noise distribution can modify the scaling
exponents, see e.g.\ \cite{alb,krug,meakin}. A somewhat 
trivial one is considering 
conserved noise (as in model B \cite{hh} for conserved dynamics of an order 
parameter), instead of non-conserved noise, implicit above. 
More far-reaching modifications are introducing {\em (i)} power-law 
correlations in the noise \cite{medina}, or {\em (ii)} a noise amplitude 
which is itself power-law distributed \cite{zhang}. This type of properties
leads to scaling exponents that depend {\em continuously} on the parameters 
that characterize the noise properties.

\item Non-locality: Another way to obtain scaling exponents depending 
continuously on model parameters is by considering long range effects on the 
surface, as represented in the dynamic equation by integral kernels that 
couple the height field values at all points on the surface \cite{non-loc}.
The parameters mentioned characterize in this case the spatial decay 
(typically as a power law) of such integral kernels. 
If non-locality appears as a consequence of a diffusing field (e.g.\ 
concentration of depositing species in CVD) being coupled to the height
field, then frequently scaling exponents are also different from those
of the four universality classes summarized above \cite{mario,guo,zhao}. 

\item Quenched vs time dependent noise: Another issue obscuring the 
identification of universality classes is the fact that two surfaces may
share the same values of the exponents $\alpha$ and $z$, but still 
be physically rather different, the noise distribution being quenched
in one case and time dependent in the other. Such a situation occurs
e.g.\ when comparing (in $d_E=2$) an EW type equation with a 
diffusion coefficient which is a random variable with a quenched 
columnar distribution \cite{lr}, with the linear MBE equation.\footnote{As 
will be seen in Secs.\ \ref{sec_anom} and \ref{sec_morpho}, in this 
example the value of the roughness exponent measured by the height-difference
correlation function is different for the first ($\alpha_{\rm loc} = 1/2$)
and second ($\alpha_{\rm loc} = 1$) examples mentioned.}
Incidentally, for surfaces growing in a (quenched) disordered medium, 
like a Hele-Shaw cell with randomly varying gap spacing, scaling exponents 
can also depend continuously on parameters characterizing the disorder 
\cite{hs}.

\end{itemize}

As can be concluded from the previous paragraphs, the values of the 
critical exponents $\alpha$ and $z$ may not suffice in order to identify the
universality class a given system belongs to. Moreover, as will be
seen below (Sec.\ \ref{sec_anom}), the most general form of the dynamic
scaling Ansatz known to date does depend on at least an additional
independent ``roughness'' exponent. There may be even accidental
or coincidental equalities among exponent sets, as occurs \cite{rothman} 
between those characterizing the KPZ equation in $d_E=1$ ---a far from 
equilibrium system--- and the ones obtained for the {\em equilibrium} 
roughening of the interface between two immiscible fluids. 
The existence of a continuum of exponent sets naturally 
questions the use of the concept of universality classes, in particular
from the practical point of view. Another inadequacy of the universality 
concept arises when interpreting
the scaling properties of some theoretical discrete models of MBE growth, 
such as the WV or DT models, which seem to belong to different universality
classes by changing the value of $d_E$ \cite{dspt}. For instance the WV model 
for instantaneous surface diffusion is in the EW class for $d_E=2$, whereas 
for higher dimensions it leads to unstable growth.\footnote{Actually, these 
phenomena point out the increased complexity of
non-equilibrium discrete models as compared to equilibrium ones, in the sense
that slight modification of the {\em dynamical} rules lead to dramatic 
changes in the asymptotic properties.} The existence of universality has even 
been questioned on theoretical grounds in the context of the very same
KPZ class \cite{newman}. Although the numerical instabilities 
adduced have been overcome and shown to be due to inappropriate schemes 
\cite{giacometti}, the strong coupling properties of the KPZ class still 
undoubtedly remain to be properly understood \cite{frey}.

\subsection{The r{\^ o}le of instabilities} \label{sec_instab}

An issue that has already arisen in the previous section is the existence
of instabilities in the context of models, or experiments, of kinetic 
roughening. Already since the first systematic studies of kinetic roughening,
physical instabilities have been well known to take place in the systems 
studied. The paradigmatic example in this context is the {\em Mullins-Sekerka}
instability arising in crystal growth from an undercooled melt and in other
growth systems (see reviews in e.g.\ \cite{books}). More recently they have 
been seen to play a key r{\^ o}le in the dynamics e.g.\ of crystal growth
by atomic or molecular beams \cite{villain}. In these systems, the 
{\em Ehrlich-Schwoebel} effect hindering the crossing of steps by adatoms
leads ---in the case of growth onto a high symmetry surface--- to development
of mounds or, when growth is on a stepped surface, to step meandering possibly
also leading to mound formation. However, in spite of their ubiquity, 
the effect of physical instabilities on the properties of the surface 
roughness has been overlooked to a large extent. Recently, however,
instabilities have been put forward \cite{mario} as a possible explanation 
for the difficulty in observing experimentally the ``ideal'' universality 
classes discussed in Sec.\ \ref{sec_ideal}. A paradigmatic example is provided 
by the (noisy) Kuramoto-Sivashinsky (KS) equation  
\begin{equation}
\frac{\partial h}{\partial t} = - |\nu| \nabla^2 h - |B| \nabla^4 h 
+ \frac{\lambda}{2} (\nabla h)^2
+ \eta .
\label{nks}
\end{equation}
This equation has been actually {\em derived} from constitutive
equations ---not merely based on scaling and/or symmetry arguments---
as a description of kinetically roughened surfaces arising
in a number of physical contexts. To name a few of them, we can mention
directional solidification of dilute binary alloys \cite{novick},
solidification of a pure substance at large undercoolings including
interface kinetics (see \cite{misbah} and references therein), erosion 
by ion beam sputtering (IBS) \cite{us_ibs}, dynamics of steps on 
vicinal surfaces under MBE 
conditions \cite{pierre-louis}, or growth by CVD or ECD \cite{mario}. 
In our context, a distinguishing feature of this continuum model 
\cite{manneville} is the {\em negative}, and therefore
linearly unstable, sign of the first term on the right hand side of Eq.\ 
(\ref{nks}). Competition between this term and the {\em stable} biharmonic
one leads to the existence of a finite band of linearly unstable Fourier
modes in the surface. The Fourier mode with the fastest increasing
amplitude leads to the production of a pattern whose appearance in principle
breaks scale invariance. The highly non-linear dynamics of the system 
eventually leads to a disordered and rough morphology whose statistics
are provided, at least for $d_E=2$, by the KPZ universality class. 
Note, incidentally, that Eqs.\ (\ref{kpz}) and (\ref{nks}) are 
{\em indistinguishable} from the point of view of symmetries, hence 
``derivations'' based on the latter typically end up by proposing the KPZ 
equation as a continuum description even in systems where the KS equation 
is a better description, due to the existence of {\em physical} instabilities,
see e.g.\ \cite{bruinsma} for the case of erosion by IBS.
Note that the crossover from the pattern-formation transient to the
asymptotic kinetic roughening can be exceedingly large in the KS system
\cite{sneppen}. Hence, it is conceivable that when studying experimentally 
systems described by the KS equation and whose asymptotic state is
effectively unaccessible, fits to some effective (and improper) exponents
are attempted that lead to incorrect conclusions on the scaling properties 
of the system. At any rate, the long crossover can even hinder the actual
experimental observation of the asymptotic KPZ scaling in systems described
by Eq.\ (\ref{nks}). Analogous conclusions can be drawn for systems in which 
physical instabilities are expected, even if the relevant dispersion
relation\footnote{Defined as the wave-vector dependent rate $\omega(k)$ 
characterizing, within a linear stability analysis, the exponential 
growth/decay of the $k$-th Fourier mode of the height field, $h_k(t) = 
{\rm const.} \, e^{\omega(k) \, t}$.} differs from the KS one, e.g.\ if it is, 
rather, provided by the Mullins-Sekerka dispersion relation, ubiquitous 
for systems with diffusional instabilities and fast interface kinetics
\cite{mario}. An open avenue for research in this connection is undoubtedly 
the interplay of noise and instabilities and its implications for the 
interface morphology, and the possible formulation of generic continuum
descriptions incorporating these phenomena, see Sec.\ \ref{sec_univ}. 

\subsection{Dynamic scaling {\em Ans\"atze}} \label{sec_anom}

As mentioned in the Introduction, one of the main problems with the
large variety of exponents found experimentally is that, in the
absence of a detailed study of the system considered,\footnote{Say, 
of a derivation of the interface equation from constitutive equations, as 
available e.g.\ in the examples mentioned in Sec.\ \ref{sec_instab}.} which
might not necessarily be available, there is no {\em a priori} argument  
allowing to identify a clear physical mechanism responsible for the
experimental data. Moreover, many of the experiments seemed to contradict 
the assumed generality of KPZ scaling, the scaling behavior actually
observed being termed {\em anomalous}. In retrospect, this induced a large 
advance in the field, at least from the theoretical point of view,
in the search for potential origins for such violation of the generic 
expectation. Thus, for instance quenched disorder was advocated as a source 
for anomalous scaling, as were various other properties of the interface
noise distribution \cite{alb,krug,meakin}. 
The problem with these was that, again, for given
experiments, no {\em a priori} arguments were provided supporting 
those kinds of properties in the appropriate noise distribution.

A potentially more fundamental reason for the spread in the values of
the scaling exponents was put forward in \cite{lr}, generalizing
results mainly obtained in the context of discrete models of surface
growth by MBE \cite{sswsp}, namely, the FV Ansatz is not the most general 
scaling behavior for a surface displaying kinetic roughening. 
Thus, it might be the case that, at least in some systems (see an explicit 
example in Sec.\ \ref{sec_example}), the values attributed
to the scaling exponents were not correct but, rather, originated in
trying to fit the data using the wrong scaling Ansatz. This type of
scaling has also been termed {\em anomalous}. At least in many cases
studied, it originates in slow dynamics of the surface {\em slopes},
which do {\em not} reach a stationary state simultaneously with the
surface height \cite{juanma}.  In its most general form to date
\cite{unican}, the dynamic scaling Ansatz is formulated in terms of
the power spectral density (PSD) or surface structure factor $S(k,t)=
\langle \hat{h}(k,t) \hat{h}(-k,t) \rangle$, with $\hat{h}(k,t)$ being
the $k$-th Fourier mode of the surface height deviation around its
spatial average for a given time $t$. Thus, for many rough interfaces
one has 
\begin{equation}
S(k,t) = k^{-(2 \alpha + d)} \, s(k t^{1/z}), \quad {\rm where} \quad
s(x) \sim \left\{ \begin{array}{lcl}
x^{2(\alpha - \alpha_{\rm s})} & {\rm for} & x \gg 1 \\
x^{2\alpha + d} & {\rm for} & x \ll 1
\end{array} \right. 
\label{psd_gral}
\end{equation}
Here, $\alpha_{\rm s}$ is an additional exponent, whose values induce
different behaviors. Namely, for $\alpha_{\rm s} < 1$ one may have FV
scaling [i.e., that expected na\"{\i}vely for $S(k,t)$ directly from the FV
Ansatz for $W(t)$, and which occurs for the ``ideal'' classes (1) and 
(2a,c) of Sec.\ \ref{sec_ideal}] or {\em intrinsic} anomalous scaling, if
$\alpha_{\rm s}$ equals $\alpha$ or not, respectively. On the other
hand, if $\alpha_{\rm s} > 1$ anomalous scaling ensues which can be
either trivial if $\alpha_{\rm s} = \alpha$, in the sense that it is
induced by the super-roughness of the interface (this is the case for the 
linear MBE equation mentioned in Sec.\ \ref{sec_ideal}), or non-trivial if
$\alpha_{\rm s} \neq \alpha$. This latter case corresponds also to
faceted surfaces featuring large values of the local slopes
\cite{unican}.  In general, signatures of anomalous scaling are
different effective roughness exponents for small and large length-scales 
---see an explicit example in Sec.\ \ref{sec_exp} below---, and finite-size 
effects in $S(k,t)$.\footnote{Inexistent within the standard FV scaling.}
To date, there is no general argument as to what kind of scaling field 
controls the value of the exponent $\alpha_{\rm s}$ in those cases in which 
it is actually independent from $\alpha$, $z$ and geometrical constants. 
Note that in these cases the number of exponents characterizing the growth
process increases as compared with the traditional assumption reviewed
in Sec.\ \ref{sec_ideal}.
There has been progress, nevertheless, in
relating intrinsic anomalous scaling with L\'evy statistics for the
local height difference distribution \cite{joo}. Moreover, in some
examples such as interfaces generated in invasion percolation models,
the value of $\alpha_{\rm s}$ can be related with the fractal
properties of the underlying cluster. This may provide a physical
interpretation for the anomalous scaling properties of the interfaces,
when these are seen to occur. Examples of experimental reports on
anomalous scaling (see also Sec.\ \ref{sec_exp} below) 
include wood fracture \cite{wood}, polymer film
growth by vapor deposition \cite{zhao}, and electrochemical or
electroless production of Cu films \cite{ecd1}. However, note these
include realizations of all types of anomalous scaling {\em but} for
the faceted surfaces characterized by $1 < \alpha_{\rm s} \neq
\alpha$.

\subsection{Fractal properties}

As mentioned above, kinetically roughened surfaces are but a specific
example of statistical {\em self-affine fractals}. A phenomenon which
has not been completely addressed, either from the theoretical, or
from the experimental points of view is the full relationship between
the geometrical fractal properties of the interface and the form of
dynamic scaling Ansatz it follows, and the scaling relations among the
critical exponents characterizing both.  

\subsubsection{Multiscaling} \label{sec_multi}

In the case of the simple FV Ansatz, there is a direct connection 
between the unique roughness exponent characterizing the rough surface
and the fractal dimension. However, it
becomes less evident for the case of anomalous scaling. As a relevant
example, for the discrete DT model \cite{dt} of MBE growth,
anomalous scaling has been seen to occur
accompanied by {\em multiscaling} properties \cite{krug2}.  These
relate to the fact that different moments of the height-difference
correlation function $G_q(r,t) = \langle |h(\mathbf{r}_1,t) -
h(\mathbf{r}_2,t)|^q \rangle$, where $r =
|\mathbf{r}_1-\mathbf{r}_2|$, scale with moment-dependent
exponents,\footnote{In order to compare with other common notations,
one has to replace $(\alpha_q,\gamma_q) \to (\chi_q,\alpha_q)$
\cite{joo}, or $(\alpha_q,\gamma_q) \to (\zeta_q,\alpha_q)$
\cite{krug2}.} $G_q^{1/q}(r,t) \sim \xi(t)^{\gamma_q} \, r^{\alpha_q}
f_q(r/\xi(t))$, with $f_q(x)$ appropriate scaling functions, and the exponents
$\gamma_q$ being related to the local height difference distribution
properties. These results for the DT model have been recently generalized
for any surface with a local height difference distribution of the
L\'evy type \cite{joo}. 
However, the general interplay between anomalous scaling
and multiscaling properties of surfaces is not known. It would be 
highly desirable to have available sufficient and/or necessary 
conditions for the occurrence of one of these phenomena as a function
of the other.\footnote{For instance, examples are known of models which display 
intrinsic anomalous scaling, but not multiscaling \cite{lr}.}
This would allow for a consistent analysis of experimental
data by, say, evaluating different moments of the height difference
correlation function in real space, and cross-checking such information
with the scaling behavior of the PSD in the form (\ref{psd_gral}).

\subsubsection{Other geometries} \label{sec_geom}

There are two other important aspects ---of at least a methodological 
nature--- which have not been addressed
in full in a systematic way, and whose study would undoubtedly clarify
the interpretation of many experimental observations, as well as
their comparison with theoretical models. Both relate to interface 
growth phenomena in geometries which differ from the strip or slab geometry.

Thus, as a {\em first} instance we have growth on circular 
($d_E=2$) or spherical geometries ($d_E=3$). Although the earliest 
systematic studies of kinetic roughening actually originate in 
the dynamics of the active zone in ``circular'' DLA or Eden clusters,
the vast majority of analytical studies have focused on the strip
($d_E=2$) or ``slab'' ($d_E=3$) geometries. Some exceptions are 
partial studies of the KPZ equation in radial geometry \cite{batchelor}.
Although, admittedly, these geometries are infrequent in applications
to thin solid film production, the clarification on the correct scaling
Ansatz to be performed in circular geometries would undoubtedly 
complete the theoretical framework. For instance, in these systems an 
essential feature is the fact that the system size is {\em not fixed} 
to a constant value. One can account for the experimental data by the 
Ansatz (\ref{psd_gral}), provided the replacement $k \to k/f(t)$ 
is performed, where $f(t) \sim t^a$ ($a$ being a constant) is a dilation 
factor taking into account the increase of the system size with growth time 
\cite{buceta}. However, differing values of $a$ are needed for different
systems, such as e.g.\ tumor \cite{bru} or plant calli growth
\cite{buceta}, being unrelated, moreover, with the value of $z$ in the
various systems. Note a possibly related non-stationarity phenomenon 
in the behavior of the local slope for the case of anomalous scaling.
It would be interesting to provide an estimation for exponent $a$ based on first 
principles, as well as deriving its value for the main interface growth 
equations, say, the EW and KPZ equations in non-conserved growth, and
the linear and non-linear MBE equations for conserved growth, see 
Sec.\ \ref{sec_ideal}.

The {\em second} issue mentioned is the deduction of scaling properties
of fully two-dimensional surfaces ($d_E=3$) from data for one-dimensional 
cuts ($d_E=2$) across them. This is a rather frequent procedure, specially
for Transmission Electron Microscopy (TEM) analysis of the
growth of multilayer films. In this case very thin cross-sections (i.e.\
one-dimensional profiles) of the multilayers are analyzed under the framework
of dynamic scaling. Thus, values for the scaling exponents are obtained.
However, discrepancies arise in the literature when these values are
compared to models in $d_E =2$ \cite{jacobo} or in $d_E=3$ \cite{elmiller}.
Although in the isotropic case the correct relationship between roughness 
exponents is clear, it becomes less immediate \cite{javi} when there exists
physical anisotropies on the substrate plane. This issue can be important 
because it also affects the correct procedure to analyze {\em anisotropic} 
growth or erosion processes. Physical realizations of such anisotropies are 
provided e.g.\ by vapor deposition of gold films \cite{roberto}
or IBS of highly pyrolytic graphite \cite{habenicht}, both under 
oblique incidence conditions (for attaching or bombarding particles, 
respectively).

\subsection{Universal (continuum) descriptions} \label{sec_univ}

As implicit from the previous sections, the physics of surface/interface 
dynamics is naturally more involved than the mere intermediate or asymptotic 
scaling laws and, even if one is interested in these (as encoding the kinetic 
roughening properties of the system), one needs take into account the 
various physical mechanisms acting on the surface morphology, which actually
may have an impact on the behavior of, say, the surface roughness.
In practice, this requires resorting to an underlying physical model of the 
phenomenon at hand in order to derive the pertinent interface 
equation\footnote{Typically through a weakly non-linear analysis.} in a 
consistent fashion. Technically, such physical models usually have the shape 
of a free or moving boundary system\footnote{See e.g.\ the chapter by 
R. Gonz\'alez-Cinca {\em et al.} in this book.} 
into which fluctuations are incorporated. This is the case in the examples
quoted in Sec.\ \ref{sec_instab} \cite{pierre-louis,mario,misbah,us_ibs,eme}.
This procedure, moreover, allows one to relate the coefficients appearing
in the Langevin equation for the interface ---say, $\nu$, $B$, and $\lambda$
in (\ref{nks})--- to phenomenological parameters (diffusivities,
temperature, mass-transfer coefficients, etc.). Admittedly, the moving
boundary problem couples effects at the interface which are strictly kinetic, 
with others of a different nature, such as e.g.\ 
mechanisms for the transport of matter, latent heat, or interfacial
free energies. Our point here is that these have to be taken into
account if one wishes to have a physical theory of kinetically roughened
interfaces which is relevant to experiments. Interestingly, one can still put
forward {\em generic} ---rather than {\em universal}, borrowing
terminology from the field of {\em pattern formation} and non-linear
science \cite{manneville}--- descriptions of kinetically roughened surfaces
that strongly follow those available for standard\footnote{In the sense of
{\em deterministic} partial differential equations.} pattern forming systems.
For instance, the deterministic stabilized KS equation has been shown
\cite{misbah2} to be generic for systems featuring bifurcations with a
vanishing wave number, provided certain symmetries occur. The advantage
of this approach is that one can provide continuum models that feature
both instabilities leading to pattern formation, together with fluctuation
effects that tend to disorder the interface and induce scale invariance
properties. The behavior observed will be then a matter of a delicate balance
between these opposing effects. Continuing with the same example, within
this approach the noisy KS equation is more generic than the KPZ equation,
in the sense that it can account for more complex dynamics that applies
to {\em both} asymptotic and pre-asymptotic features. It will be of great 
interest if, proceeding along these lines, one could provide generic continuum
descriptions that feature for instance anomalous scaling, known e.g.\ in the 
context of ECD to be a preasymptotic feature \cite{mario_ecd}.

\section{Status of the experiments} \label{sec_exp}

As mentioned above, regarding the experimental issues we will mainly focus 
our attention on processes related to the growth or erosion of the 
surfaces of thin solid films. These processes take place usually under 
far-from-equilibrium conditions. Several techniques have been employed to
characterize the growth morphologies appearing; among them 
Scanning Tunneling Microscopy (STM), Atomic Force Microscopy (AFM), 
X-ray reflectivity, Reflection
High-Energy Electron Diffraction (RHEED) and Transmission Electron
Microscopy (TEM). The suitability of each of these techniques depends on the
specific sample to be studied, the length scale to be sampled, the
interface roughness, and the character of the study to be made, since
not all of them yield values for all the scaling exponents \cite{kp}.
Besides, for instance X-ray reflectivity measurements can be carried out 
{\em in situ}, whereas Scanning Probe Microscopy (SPM) has to be employed 
{\em ex situ}. Nevertheless, in later years SPM techniques seem to be 
predominant for analyzing surfaces of thin solid films, since they allow for a wide 
window length for analysis (from nanometers up to several tens of microns) 
and they do not require any special sample preparation (in particular with 
AFM). They allow to measure the surface morphology for different growth 
(erosion) times and thus directly obtain e.g.\ the value of exponent
$\beta$. Moreover, from the analysis of the topography one can obtain 
different observables, such as the power spectral density (PSD), the 
height-height correlation function, the height autocovariance, etc.,
from which the values of other critical exponents, such as $\alpha$ and 
$z$ can be determined. However, it has been pointed out that SPM techniques can
overestimate the value of $\alpha$ when this is relatively small
\cite{aue, mannel}, although consistent values have been also found
when the AFM values have been compared to those obtained by different
techniques for a given system \cite{souza,mcrae,dharma,durr,smets,stromme}. 
In the following sections we consider additional difficulties that can be met 
in experimental works.

\subsection{Scaling ranges}

In principle, in order to characterize appropriately the growth process
under study,\footnote{Without {\em a priori} assumptions on the universality
class.} one should determine all independent exponents by separate means, 
making no previous assumptions on scaling relations of the type 
$\beta$ = $\alpha/z$, that can fail due e.g.\ to specific initial 
conditions employed \cite{rough}. In this sense it is very
useful to identify the characteristic lateral correlation length with
some morphological film surface feature (for instance, with the grain
structure). Also, a relevant issue concerns the magnitude of the
window (either spatial or temporal) over which the value of the
exponent is being determined. This is important since theoretical
models usually deal with windows of several decades, whose extent can be 
very hard to achieve in experimental systems \cite{malcai}. As a
consequence of this limitation, it is very convenient to carry
out deposition (erosion) experiments for very long times in order to
be able to analyze the truly asymptotic growth (erosion) scaling
behavior \cite{us-cvd,us-dots}. Admittedly, the relevant time scales
may lie beyond the physical constraints of the experimental setup. 
Finally, another experimental problem found in many systems \cite{us-cvd} 
is related to the existence of {\em crossovers} between different growth 
regimes. This limitation is aggravated by the fact that the crossovers are 
in practice not sharp but, rather, even more gradual than as seen in many 
theoretical models. Thus, reliable identification of the extent of the 
different growth regimes becomes hampered, their temporal or spatial extension
being reduced.

\subsection{Morphological analysis of experimental data} \label{sec_morpho}

Any rationalization of experimental data draws unavoidably upon some
theoretical model, and can thus be biased by the limitations of the 
latter. This fact also occurs in our context, where perhaps the best 
example is the existence of the anomalous scaling. Note that
the theoretical understanding of the peculiarities it introduces with respect to 
the simplest FV scaling Ansatz has been taking place during the last ten 
years, well after formulation of the latter \cite{fv}.
Specifically, as was explained above, in principle several different 
types of dynamic scaling can be distinguished: {\em (i)} the simplest
one is the classical FV scaling, that is associated
with overlapping PSD [the function $S(k,t)$ defined in Sec.\ \ref{sec_anom}] 
or height-difference correlation [the function $G_2(r,t)$ introduced
in Sec.\ \ref{sec_multi}] functions for different deposition 
times,\footnote{For length-scales smaller than the lateral correlation length.}
and with only two independent exponents, $\alpha_{\rm s} = \alpha$ and $z$, 
with $\alpha_{\rm s}$ the spectral exponent defined in Eq.\ (\ref{psd_gral}). 
{\em (ii)} The merely ``super rough" case in which, while the PSD 
graphs overlap for different times, the height-difference graphs do not, 
but are shifted upwards as deposition (erosion) time increases. 
In this case one has to distinguish between local ($\alpha_{\rm loc} \equiv 1$)
and global ($\alpha > 1$) roughness exponents,\footnote{The local roughness
exponent $\alpha_{\rm loc}$ coincides with exponent $\alpha_2$ as 
defined in Sec.\ \ref{sec_multi}. Equivalently, it characterizes the spatial
scaling of the local width $w^2(\ell,t) = (1/\ell^d) \langle \sum 
(h(\mathbf{r},t) - \bar{h})^2 \rangle$ where spatial averages are restricted 
to boxes of lateral size $\ell$. Thus, for $\ell \ll t^{1/z}$ one has 
$w(\ell,t) \sim \ell^{\alpha_{\rm loc}}$.} since they appear different
as measured from $G_2(r,t)$ or from $S(k,t)$, which are sensitive to,
respectively, local or global fluctuations. In this case $\alpha_{\rm s} = 
\alpha > 1$, and the value of the local roughness exponent is fixed to 1 
by geometrical constraints \cite{meakin}, hence it is not really an 
independent exponent. {\em (iii)} The ``intrinsic" anomalous case in which 
neither the PSD nor the height-difference correlations overlap for different
times but are, rather, shifted as deposition (erosion) time increases
[for some faceted surfaces $\alpha_{\rm s} > 1$, which induces time shift
only in the PSD and not in the $G_2(r,t)$ function].
In this case, again local and global fluctuations are accounted for by
different roughness exponents, there existing now {\em three} really
independent exponents. 

It is clear from the above that proper analysis of the data requires 
previous inspection of the two functions, $G_2(r,t)$ and $S(k,t)$, in order 
to anticipate the scaling behavior at least in terms of the existence of the
mentioned non-overlapping\footnote{As mentioned in Sec.\ \ref{sec_anom},
for a finite system the non-overlapping features of the PSD amount at 
saturation to finite size effects, inexistent for FV scaling.} 
features. Once it is identified, the adequate procedure of
analysis of the data can be found in the literature. In general,  
the most elegant method consists in the collapse of the height-difference
correlation or/and the PSD experimental curves, according to the
theoretical scaling relations \cite{smets}. Another possibility
requires performing roughness analysis at short and long length scales in
order to measure the local and global roughness exponents 
\cite{ecd1,saitou,yang}. 

Different situations can be found when one examines the existing literature 
regarding this subject. First, there are many studies in which the authors 
did not realize that anomalous scaling was present in their systems, or they 
did but a consistent analysis was not performed because, rather, FV scaling 
was assumed \cite{lee,zhao,reinker,bray,liu}. An example of this can be found
in an earlier paper by one of the present authors \cite{aplcvdluis}, 
see details in Sec.\ \ref{sec_example} next. A second set of 
experiments exists in which the authors explicitly identified the anomalous 
behavior and analyzed it accordingly, mostly through the behavior of
the interface width for different length scales \cite{jeff,ecd1,saitou}. 
It should be noted that most of these studies in three-dimensional systems 
are relatively recent, and are mostly related to electrodeposition systems. 
In general, we can say that the description of anomalous scaling is an 
example of a situation in which the theoretical developments are ahead from 
the experimental analysis. There is still, however, an important issue than 
remains unsettled, related with the physical origin of this type of behavior. 
While the theoretical papers on the subject address more mathematical aspects, 
one can already obtain some light on this issue from the (scarce) experimental
works on the matter. As we have mentioned, in the context of thin solid film growth,
most of these are focused mainly on electrodeposited films, suggesting that 
non local ---in this case bulk diffusion--- effects are likely responsible 
for such behavior. This scenario would be consistent with different local 
growth rates, which is qualitatively consistent with other explanations 
suggested for two-dimensional systems of a different nature \cite{hs}. 
It should be stressed, however, that the physical origin of anomalous scaling
is an issue that still has to be explicitly addressed. Another important 
(possibly related) aspect regarding consistency of the analysis of
the experimental data, especially when instabilities are found, is
that one of multiscaling. It is important to check explicitly 
for such type of behavior in any chosen system for, in the case it 
does exist, it does not make sense to refer to a single set of scaling 
exponents but, rather, one should consider an (in principle, infinite) 
set of exponents, one for each moment of the height difference distribution. 

\subsubsection{An explicit example} \label{sec_example}

\begin{figure}
\begin{center}
\includegraphics[width=9cm]{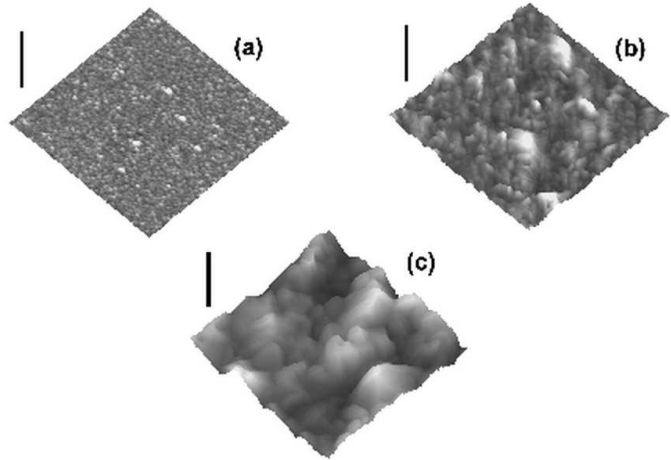}
\caption{$4 \times 4$ $\mu {\rm m}^2$ STM images of Cu films grown by
low-pressure CVD onto Si(100) substrates, for values of the thickness:
44 nm (a); 334 nm (b), and 2220 nm (c). The vertical bars represent 1000 nm.}
\label{fig_morph}
\end{center}
\end{figure}

\begin{figure}
\begin{center}
\includegraphics[width=7cm]{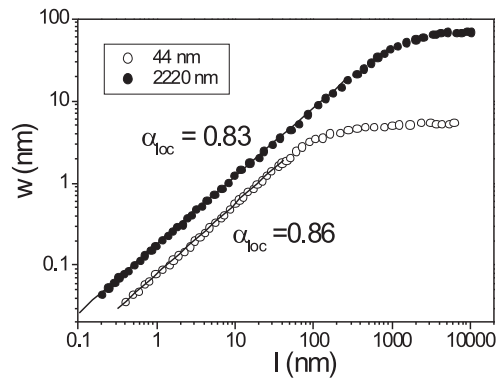}
\caption{Logarithmic plot of the local roughness, $w(\ell)$, vs window
size, $\ell$, for the 44 nm thick film (lower curve) and for the 2220 nm 
thick film (top curve). The values indicated for $\alpha_{\rm loc}$ provide 
the slopes of the corresponding solid lines. This figure corresponds to 
Fig.\ 3b of Ref.\ \cite{aplcvdluis}.}
\label{fig_local}
\end{center}
\end{figure}

\begin{figure}
\begin{center}
\includegraphics[width=6.2cm]{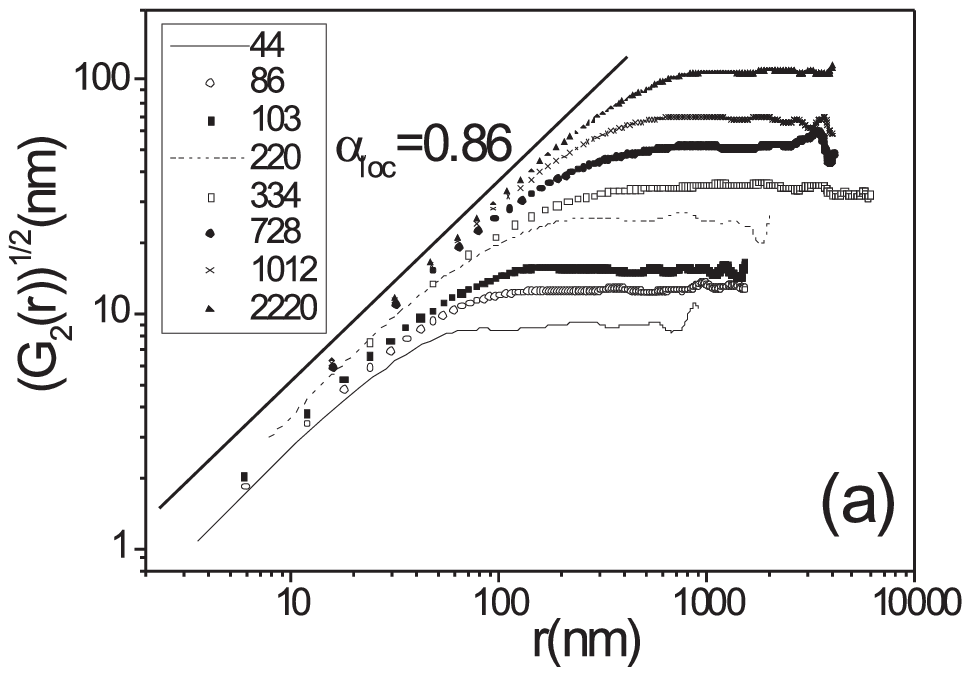} \includegraphics[width=6.2cm]{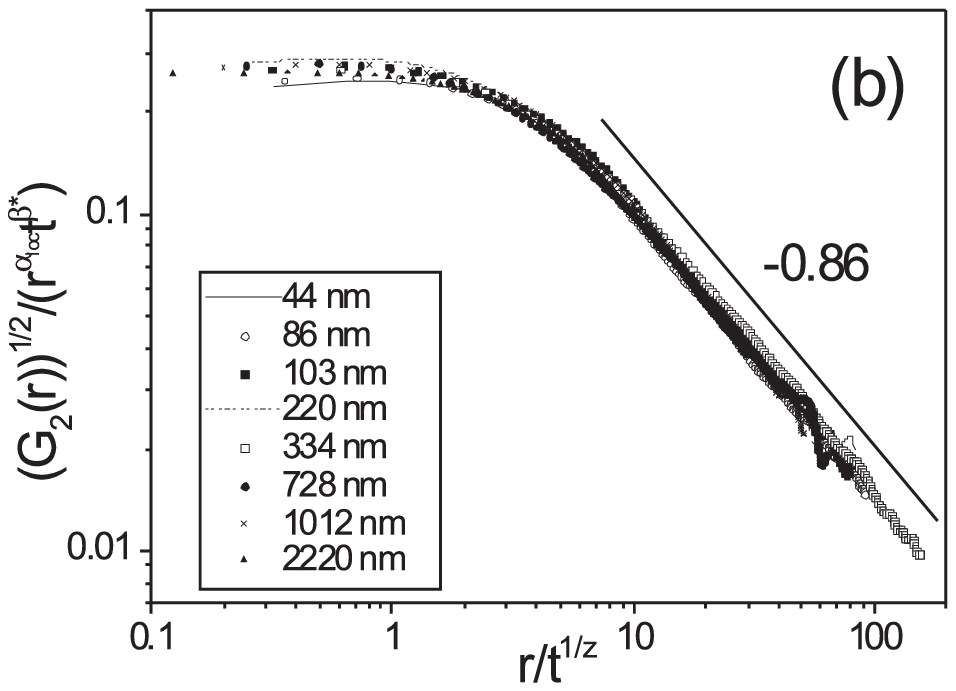}
\caption{(a) Logarithmic plots of the square root of $G_{\rm 2}(r,t)$ vs $r$ 
for different thicknesses $t$ (corresponding thickness values and symbols are
indicated in the figure). The value indicated for $\alpha_{\rm loc}$ provides 
the slope of the thick solid line.
(b) Collapse of the data provided in panel (a). Vertical coordinates are 
rescaled by a factor $r^{\alpha_{\rm loc}} \, t^{\beta_*}$, using 
$\alpha_{\rm loc}=0.86$ and $\beta_* = 0.11$, while horizontal coordinates 
are divided by a factor $t^{1/z}$ with $z=1.59$. For the sake of reference,
a thick solid line has been drawn whose slope equals $-0.86$.}
\label{fig_g}
\end{center}
\end{figure}

In order to provide an example that illustrates the various difficulties met 
in this type of works, we devote this section to reanalyzing an experiment
by one of the present authors \cite{aplcvdluis} in the light of the more 
recent developments on the types of dynamic scaling Ans\"atze. We will see 
that even doing so, a complete understanding of the corresponding growth 
process is still open. The experimental system is growth by CVD of copper 
films on Si(100) substrates. In Fig.\ \ref{fig_morph} we display three STM
images of films with thickness (proportional to time, given the constant
growth rate imposed) of 44 nm (a), 334 nm (b) and 2220 nm (c), respectively. 
The film roughening and coarsening are evident in the figure.
As can be observed in Figs.\ \ref{fig_local} (corresponding to Fig.\ 3(a) in 
Ref.\ \cite{aplcvdluis}) and \ref{fig_g} for the local surface roughness
$w(\ell)$ and for the height-difference correlation function $G_2(r,t)$, 
respectively, the surface morphology displays anomalous scaling in the sense 
that curves for different times do {\em not} overlap for small values of $r$.
Specifically Fig.\ \ref{fig_local} contains data for the thinnest and thickest 
copper films.
However, in Ref.\ \cite{aplcvdluis} the scaling analysis was made according to 
the FV Ansatz, with results for the critical exponents
$\alpha_{\rm loc}= 0.81 \pm 0.05$, $\beta= 0.62 \pm 0.09$, and $1/z = 0.74$,
the last value having a relatively large error bar since it was obtained by 
averaging the grain distribution functions. In order to improve the 
scaling analysis, we present now in  Fig.\ \ref{fig_g} the $G_2(r,t)$ curves 
for various film thicknesses $t$. 
The anomalous scaling displayed by $G_2(r,t)$ can be seen even
more clearly in the systematic upward shift of the different curves  
with increasing thickness. From the slope of the curves for small $r$ values
(such as was the procedure employed in \cite{aplcvdluis}) we already obtain 
$\alpha_{\rm loc} = 0.86$. In order to characterize quantitatively 
the anomalous scaling, we have used the data from Fig.\ \ref{fig_g}(a) and 
collapsed the curves [see Fig.\ \ref{fig_g}(b)] by dividing the 
horizontal axis by $t^{1/z}$, and by dividing the vertical axis by the 
product $r^{\alpha_{\rm loc}} \, t^{\beta_*}$, with $\beta_* = (\alpha - 
\alpha_{\rm loc})/z$ \cite{lr}. 
The best collapse of the data is obtained for $1/z = 0.63$ and 
$\beta_* = 0.11$. These results imply that the system is characterized by 
$\alpha = 1.03$, $\alpha_{\rm loc} = 0.86$, $z = 1.59$, $\beta = 0.65$, and 
$\beta* = 0.11$. Note that the collapse is consistent since the behavior
of the scaling function for large arguments should decay as 
$(k t^{1/z})^{-\alpha_{\rm loc}}$, as indeed occurs in Fig.\ \ref{fig_g}(b).

Although there are some discrepancies between the values of the scaling 
exponents reported in \cite{aplcvdluis} and those obtained here, the 
difference lies within the error bars. Yet, conceptually there is a large 
difference, since the system seems to display intrinsic anomalous scaling, 
and is thus characterized by three, rather than two, independent exponents. 
Any theoretical model of the present growth system has to be able to account 
for this fact. 
Moreover, from the physical point of view a result that remains to be 
explained is the large value of $\beta > 0.5$. This {\em unstable} growth 
could be due to non local effects such as shadowing, that are likely to occur 
in CVD, or alternatively to the existence of step-edge barrier effects as 
recently shown for copper films deposited by sputtering \cite{huang}. 
In summary, the growth system studied in this section certainly exemplifies 
several (possibly generic) features of experimental studies in the field.
Namely, it displays unstable growth reflected in a large growth exponent 
and in the existence of intrinsic anomalous scaling, that remained unnoted
when studying the properties of the roughness under the FV Ansatz. The 
more complex scaling uncovered might provide clues as to the 
understaning of the relevant physical mechanisms acting in the system,
and remains to be explained theoretically.

\subsection{Identification of growth modes} 

As described above, the experimental works are mainly based on
measurements of two or three of the scaling exponents and their
further comparison with those predicted by the different
existing growth models. In this analysis, one should consider the
physical properties of the growth system, such as growth temperature,
vapor pressure of the depositing material, kinetic or diffusive
control of growth, sticking coefficient, etc.\ in order to be able to
neglect those processes which are irrelevant and should not contribute, 
say, to the continuum equation for the interface, if such a description is 
available. For instance, if the growth temperature is relatively high,
surface diffusion should be taken into account, or if the vapor
pressure of the depositing material is very low the Gibbs-Thompson
effect (and thus terms accounting for surface tension) can be
neglected. In this sense, the very morphological data obtained for the
system can also contribute to including or discarding terms in the equation
of motion. This is the case for the skewness of the growing interface, 
that measures the symmetry of the interface. A negligible value indicates 
that the surface is quite symmetric with respect to the average height. In
contrast, noticeable values of the skewness, whether positive or negative, 
indicate that the surface presents large protuberances or depressions
that make the morphology assymetric with respect to the average height. 
Under such circumstances, 
the continuum equation should include terms with appropriate derivatives of the
height field \cite{palasan1,palasan2}. Moreover, the analysis of the behavior 
of the surface slope can also contribute to identifying the growth mode;
particularly in the case in which step-edge barriers are present in the
growth leading e.g.\ to mounded morphologies \cite{smilauer}. These arguments
may be successfully combined with others of the types described in Sec.\
\ref{sec_univ} in order to produce continuum descriptions for various
thin solid film growth processes. Thus, many
systems have been described and understood adequately under the
framework of the dynamic scaling theory. Among them we can mention
thin solid films grown by physical evaporation, sputtering, sedimentation,
electrodeposition and chemical vapor deposition
\cite{us-cvd,lvb,jeff,ecd1,sacedon,luis}. 
Among the cases where a clear identification of the growth mode was 
{\em not} possible we can distinguish roughly two classes. {\em First}, 
those cases in which some exponents are very close to those of an
existing model but other fail to agree 
\cite{palasan1,you,thompson,kahanda,chevrier} or where the adscription
to a given model is made without determining, at least, two of the exponents
\cite{hahn}. In this group we can find some cases in which the growth
mode seems to be close to KPZ, but it cannot be demonstrated. 
{\em Second}, systems for which the exponents simply are not explained by 
any existing model or class \cite{yoon,lu1,lu2}. In many of these cases, the 
exponent that more clearly fails to agree is the growth exponent $\beta$, which frequently 
results larger than the theoretically expected one. Usually, although 
most models of the type considered in Sec.\ \ref{sec_ideal} 
predict\footnote{For the case $d_E =3$ relevant to thin solid film 
production.} a value of $\beta$ smaller than 0.3, larger values 
are reported for many systems \cite{durr}. Even more, in  
many cases $\beta > 0.5$ is found, see Sec.\ \ref{sec_example}.
Given that $\beta=0.5$ is the value predicted by the random deposition model, 
which does not include any relaxation mechanism, this situation is 
referred to as unstable or rapid roughening in the literature 
\cite{krug,durr}. In fact, ``rapid roughening has been reported for a number 
of systems \cite{saitou2,fang,lanczycki,collins,press} but no general
mechanism has been identified", as pointed out recently by D\"urr {\em et
al.} \cite{durr}. Usually, the authors of the corresponding works
usually try to develop more or less simple models that consider some
physical parameter relevant in their systems to try to understand and
interpret their results. Thus, phenomena like surface heterogeneity
\cite{durr}, ``cluster coalescence" \cite{palasan1}, mass transport
diffusion effects \cite{kahanda}, sticking effects \cite{lu1,lu2},
grain growth \cite{lita}, step-edge barrier effects \cite{ernst}, 
and shadowing effects \cite{bellac}, among others, have been invoked and
sometimes modeled to explain the observed behaviors. 

\subsubsection{Polycrystallinity, step-edges and other issues}

In relation to some of the above effects, an important open issue has 
to do with the polycrystalline character of many of the films studied. 
In fact, many of the growth systems with wider impact on technological 
applications lead to polycrystalline films. Already in one of the earlier 
experimental reports \cite{thompson}, dated 1994, on the growth scaling 
behavior of gold films, the authors noted that ``one significant discrepancy 
between the experimental conditions and all of the deposition models is that 
grain boundaries are not included in the modeling". Generally, these models
do not include any mechanism to account for the formation and
development of the polycrystalline grains. These grains are expected
for the majority of vapor-deposited and sputtered-deposited
films. Thus, it would be very important for non equilibrium growth
models to consider also growth on the surfaces of the crystallites,
which are misoriented with respect to one another \cite{thompson}. This
same idea was also considered by Kardar two years later in his short
review on dynamic scaling phenomena in growth processes
\cite{kardar}. In that work Kardar insisted that ``variations in
crystallinity had so far been left out of the theoretical
picture". This statement is still valid. In his work Kardar suggested
the suitability to include in the models an additional order parameter
$M(\mathbf{r},t)$ describing the local degree of crystallinity.\footnote{See 
analogous theoretical work for growth of a {\em binary} film in 
\cite{drossel}.} Moreover, this generalized approach should account for 
experimental instances which have been nevertheless adequately described 
according to the existing models for polycrystalline systems \cite{jeff}. 
Thus, further research addressing
different experimental issues, such as the influence of the growth
substrate and other growth parameters, could contribute to obtaining a
deeper knowledge of the influence of polycrystallinity on film growth
dynamics. Another related issue is the investigation of the possible 
r\^ole of step-edge barriers on film
growth for polycrystalline materials. The effect of these barriers has
been invoked for describing growth for different metallic and
semiconductor systems \cite{krug,villain}. However, its r\^ole in
polycrystalline systems is more debated. It is assumed in a
polycrystal that step-edge barriers will be improbable if growth starts 
out with randomly orientated grains and grain sizes much smaller than film 
thickness \cite{jeff}. However, recently step-edge barrier effects have been 
proved to be a key ingredient on the growth of one micron thick 
polycrystalline copper films by sputtering, when the
grains have a size close to 200 nm \cite{huang}. 
Despite these efforts, the above issues exemplify a situation in which 
theoretical developments are well behind experimental
findings, there being in particular a need of further theoretical
studies that considering unstable interface growth. From
the experimental point of view, more efforts should be addressed to 
studying the dependence of the unstable growth behavior on growth
parameters. Recent examples of such type of studies can be found, for 
instance, in the report on dependence of anomalous scaling on the current 
density for copper electrodeposition \cite{ecd1}, or that on the dependence
of the growth instability on the sticking coefficient in the case of
growth of silica films by CVD from silane precursor \cite{us-cvd}. 

Finally, it would be important to check systematically for the presence of 
multiscaling, as well as for non-trivial noise correlations in unstable
experimental growth systems. This would allow for a better insight into the 
physical mechanisms behind growth instabilities and, thus, would improve
theoretical modeling of such type of systems. Regarding comparison with 
experimental data various improvements would be required in the theoretical
descriptions. Thus, most theoretical models
work within the small-slope approximation, that is assuming morphologies in
which $| \nabla h|$ is small. Obviously, this is not the general case for
unstable features. Also, theoretical analysis of growth instabilities 
requires simulations to be carried out on systems with considerably larger 
sizes, in order to probe the asymptotic properties. This fact is aggravated
by the fact that most of the thin solid film or surface erosion systems are in 
$d_E=3$ (i.e.\ onto fully bidimensional substrates) \cite{guo}. Thus, one can 
find many publications where the experimental system studied is effectively 
three-dimensional, but its scaling behavior is contrasted to that obtained 
from two-dimensional models. As the values of the scaling exponents depend on 
dimensionality of the system, this situation could limit full understanding of 
the system behavior.

\section{Conclusions and outlook}

The situation described in the previous pages clearly indicates 
a mismatch between theoretical predictions and experimental findings 
in the area of thin film surface growth. Also, it is questionable that the 
different experimental results as they stand can be adscribed to 
or understood under a reduced group of universality classes. However, 
most of the reported experimental results indeed show clearly kinetic 
roughening properties. Even more, several of them have been adequately 
adscribed to some of the former universality classes. Thus, although this 
situation has generated some confusion in the scientific community we
believe that further research in this field is still worth undertaking. Our aim
in this chapter has been to describe this situation and suggest future
lines of research. On the theoretical side, we expect that the search for
{\em generic} descriptions that capture both, the pattern-forming mechanisms
and the relevance of strong fluctuations, will provide a more accurate 
theoretical framework for the field. Progress along these lines will 
undoubtedly come through addressing specific growth systems, and after
a suitable generalization process. Naturally, benefit can be gained from 
studies of a more general nature such as, e.g.\ on the origin of anomalous
scaling and of multiscaling in fluctuating interfaces. Note that
many of the experimental systems whose behavior does not lie within any of 
the former universality classes do show kinetic roughening with the
features of unstable or rapid roughening growth, and/or with anomalous 
scaling. Thus, we have tried to bring the attention of the reader, either 
experimentalist or theoretician, to some topics that are not a particular
focus in the literature (anomalous scaling, instabilities, polycrystallinity, 
etc.), with the aim that in future their consideration and study may
contribute to reaching a deeper knowledge of the phenomenon of kinetic 
roughening. In this sense, more systematic and rigorous experimental studies addressing 
the influence of the different growth parameters on the kinetic roughening 
properties would be welcome.

\section*{Acknowledgments}

The authors wish to acknowledge J.\ M.\ Albella, T.\ Ala-Nissila, 
J.\ Asikainen, M.\ Auger, A.-L.\ Barab\'asi, M.\ Castro, R.\ Gago, 
I.\ Koponen, K.\ B.\ Lauritsen, J.\ M.\ L\'opez, H.\ A.\ Makse, E.\ Moro, 
F.\ Ojeda, M.\ A.\ Rodr\'{\i}guez, M.\ Rusanen, R.\ Salvarezza, 
A.\ S\'anchez, and O.\ S\'anchez for their assistance and collaboration 
during the last years in many issues related to this work. They also 
acknowledge partial support by MCyT (Spain) grants Nos.\ BFM2000-0006,
BFM2003-07749-C05-01 (R.C.), and MAT2002-04037-C03-03, BFM2003-07749-C05-02 (L.V.).

\end{document}